\documentclass[11pt]{article}
\usepackage[english]{babel}
\usepackage{amsmath} 	
\usepackage{amsfonts}
\usepackage{amsthm}
\usepackage{amssymb}

\usepackage[round]{natbib}
\usepackage[utf8]{inputenc}
\usepackage{hyperref}

\usepackage{graphicx}
\usepackage[font=small]{caption}
\usepackage{subcaption}

\usepackage{comment}
\usepackage{enumerate}
\usepackage[normalem]{ulem} 
\usepackage{chngcntr}
\counterwithout{paragraph}{subsubsection} 
\usepackage{lettrine} 
\usepackage{paralist} 
\usepackage[hmarginratio=1:1,top=32mm,columnsep=20pt]{geometry} 

\theoremstyle{plain}

\theoremstyle{definition}

\theoremstyle{remark}

\def\g{\boldsymbol}
\def\pr{{\rm Pr}}
\def\ci{\perp\!\!\!\perp}

\DeclareMathOperator*{\argmax}{arg\,max}

\begin{document}
\linespread{1.5}\selectfont

\thispagestyle{empty}
\begin{center}
{\LARGE \textbf{Modeling Website Visits}\\}
\phantom{adsf}
{\large Adrien Hitz and Robin Evans\\}
{\large \textit{University of Oxford}\\}
\end{center}

\small
\setlength{\leftskip}{1cm}
\setlength{\rightskip}{1cm}

We propose a multivariate model for the number of hits on a set of popular websites, and show it to accurately reflect the behavior recorded in a data set of Internet users in the United States. We assume that the random vector of visits is distributed according to a censored multivariate normal with marginals transformed to be discrete Pareto IV and, following the ideas of Gaussian graphical models, we enforce sparsity on the inverse covariance matrix to reduce dimensionality and to visualize the dependence structure as a graph. The model allows for an easy inclusion of covariates and is useful for comprehending the behavior of Internet users as a function of their age and gender.

\setlength{\leftskip}{0pt}
\setlength{\rightskip}{0cm}
\normalsize

\newpage
\section{Introduction}

\begin{figure}
\begin{center}    
\begin{subfigure}{0.76\textwidth} 
        \includegraphics[width=\textwidth, trim=1cm 0.6cm 1cm 1.8cm]{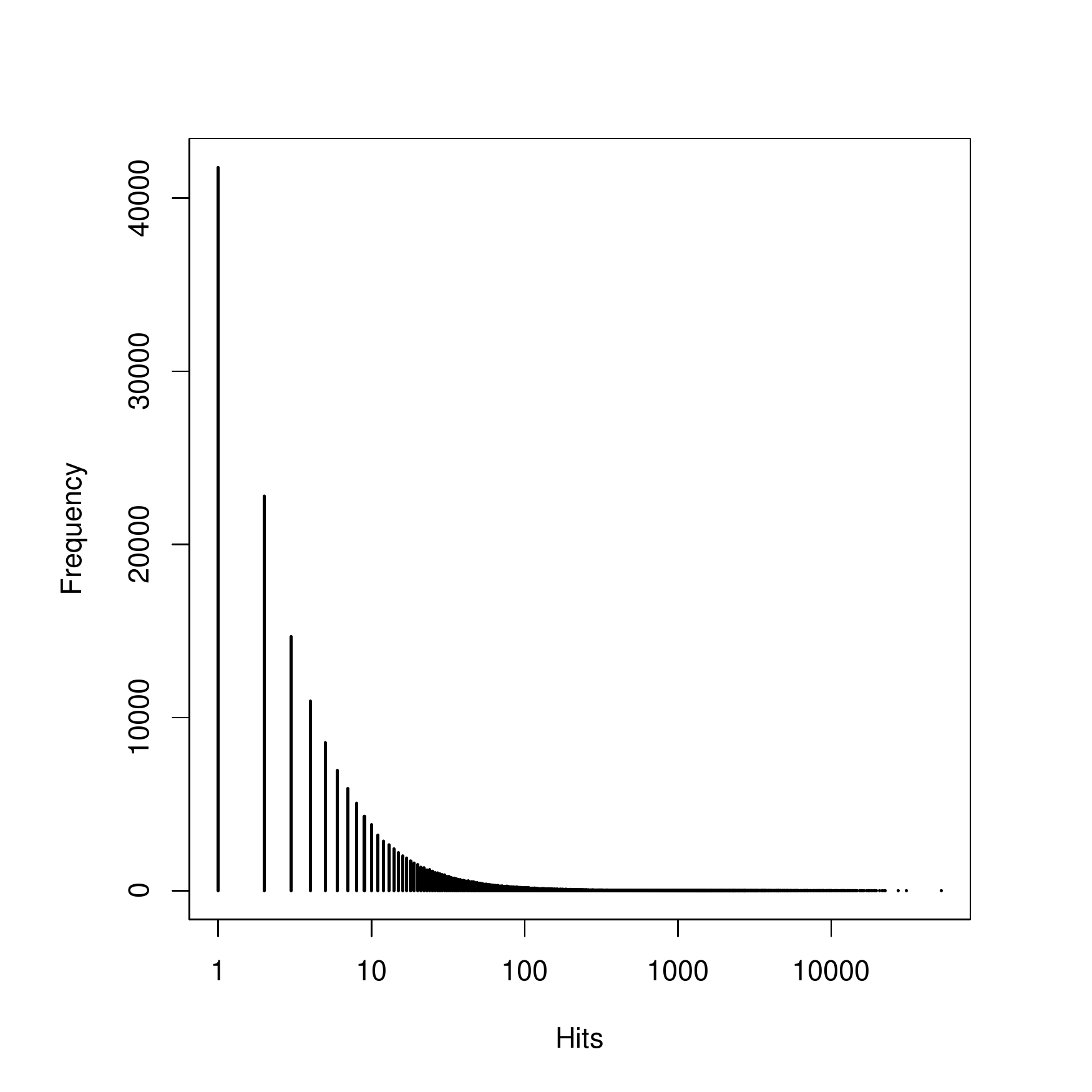}
    \end{subfigure}
    
 \begin{subfigure}{0.49\textwidth} 
        \includegraphics[width=\textwidth, trim=1.2cm 1cm 1.2cm 1cm]{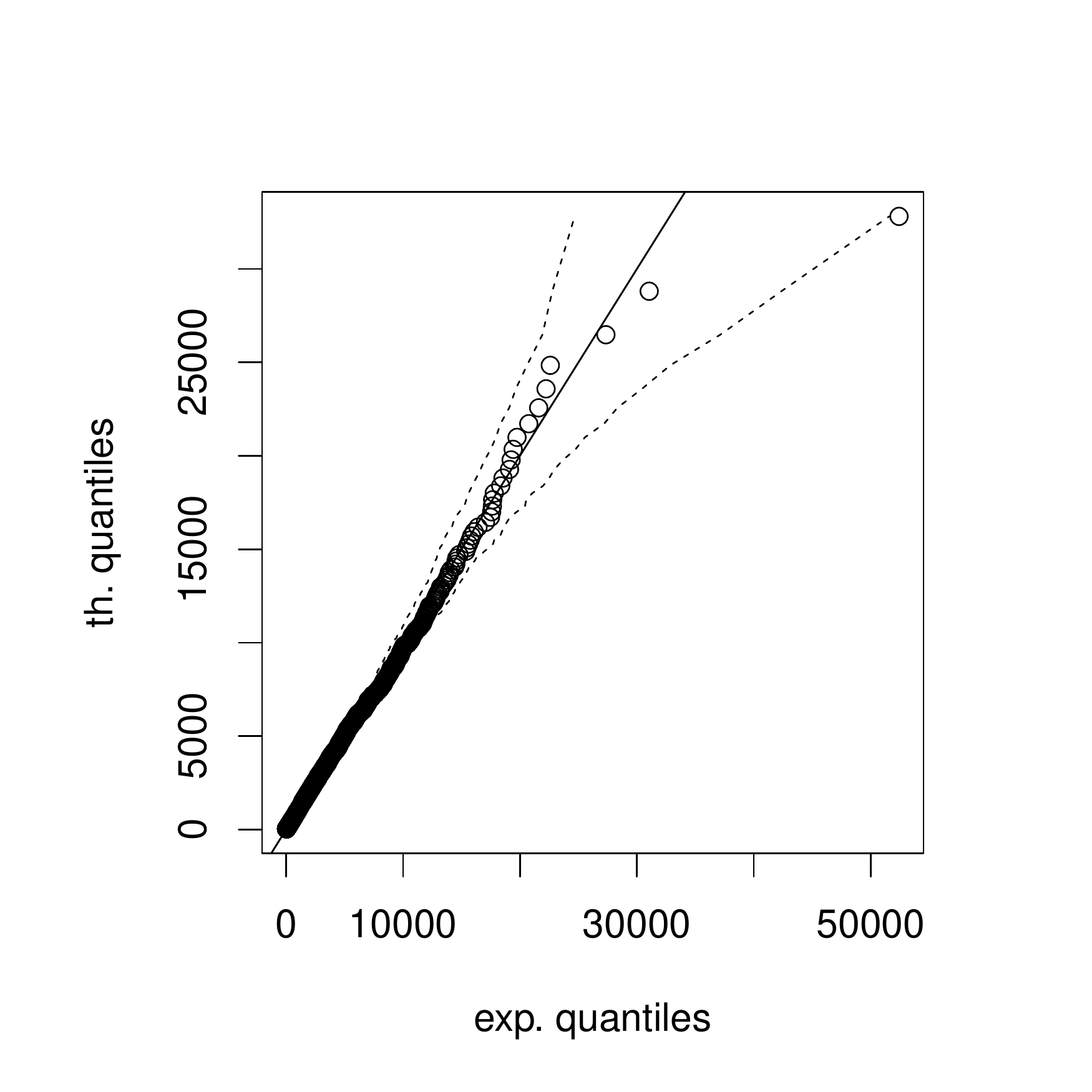}
    \end{subfigure}
\begin{subfigure}{0.49\textwidth}
        \includegraphics[width=\textwidth, trim=1.2cm 1cm 1.2cm 1cm]{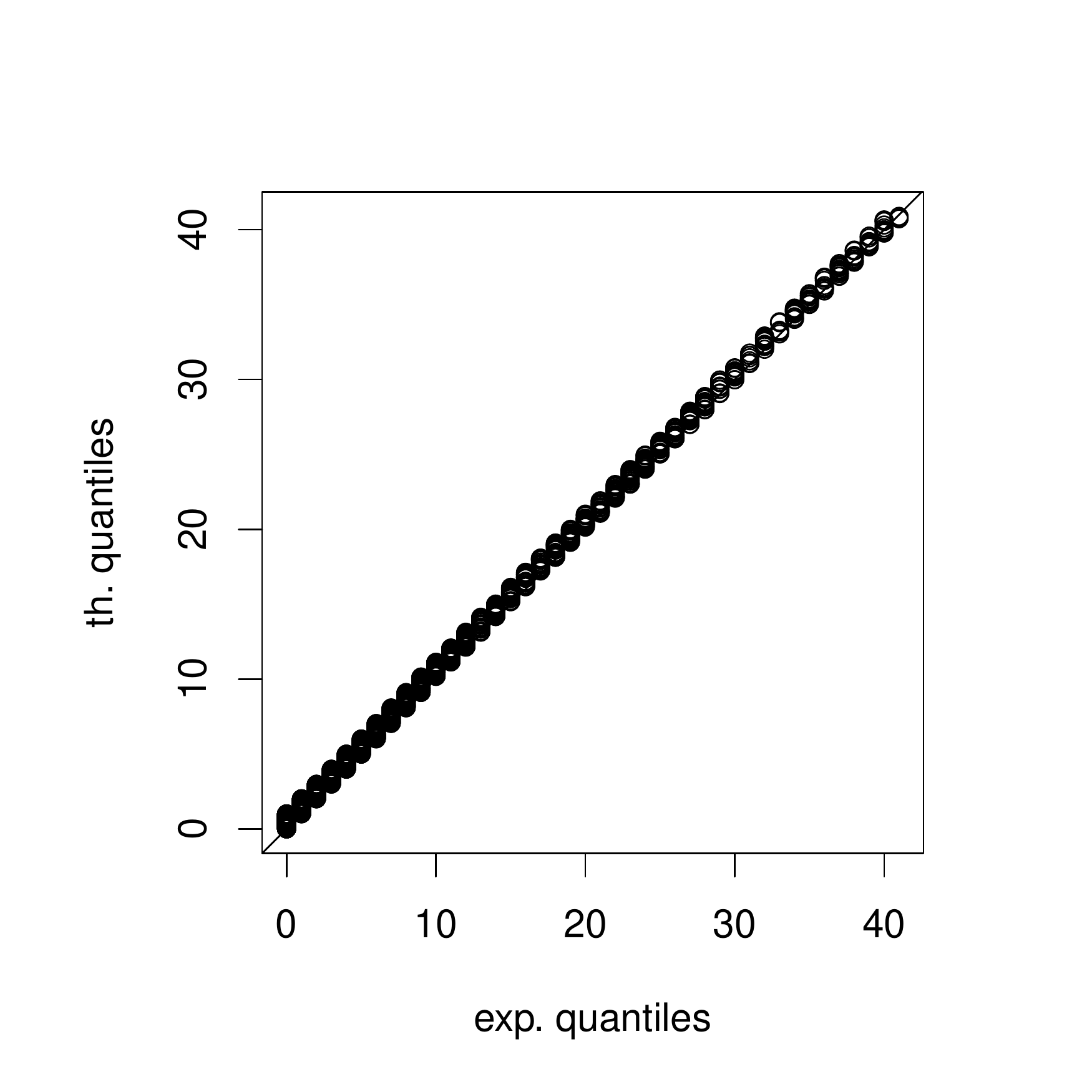}
    \end{subfigure}
               \end{center}

    \caption[Frequency table of website visits and QQ-plots]{Frequency table of about $200\, 000$ positive hits to any of the $99$ most visited websites (top), QQ-plot for a discrete Pareto IV distribution fitted to these data (bottom left) and the same plot for quantiles smaller than $0.8$ only (bottom right). The dotted lines denote $95\%$ pointwise confidence intervals.}
      \label{fig:aggrHits}
\end{figure}

Interest in the study of online behavior has grown with the importance of the Internet itself in our society, facilitated by the advent of rich data sets on browsing history; see for instance \citealp{bucklin2003model} and \citealp{johnson2004depth}. Several models have been suggested for the number of hits to multiple websites or for the number of page viewed, such as a time dependent stochastic model \citep{park2004modeling}, a multivariate generalization of the negative binomial distribution \citep{danaher2007modeling} and a Gaussian copula with negative binomial marginals \citep{danaher2011modeling}.


In this article, we consider a data set from Nielsen Holdings PLC, an information and measurement company, containing the number of visits of $19\, 436$ users to the $99$ most visited websites in the United States during one month. The table below displays hits of two users to the ten most popular websites.

\begin{center}
\small
\begin{tabular}{ l | c c  }
Website & User 1 & User 2 \\
\hline
Google  &   2 & 1\\
Yahoo!    &  155 & 668 \\ 
Facebook &  0 & 0 \\
YouTube & 0 & 2 \\
MSN/WindowsLive/Bing & 0 & 0\\
AOL Media Network & 0 & 0\\
Amazon & 0 & 0\\
Ask Search Network &  1 & 0\\
Wikipedia &0 & 0\\
 eBay &  0 & 0\\
\end{tabular}
\end{center}

We are treating the number of visits as i.i.d.\  realizations $\g x=\{\g x^{(k)}\}_{k=1}^n$ of a random vector $\g X=(X_1,\ldots,X_d)$ for $d=99$ and $n=19\, 436.$ Modeling these data seems challenging because the marginals $X_i$ have very heavy-tailed distributions exhibiting peaks at zero and are strongly dependent. We will present a simple multivariate distribution that fits the data relatively well. Before starting, we leave aside about $10\%$ of the observations and refer to them as test data.

\section{The Discrete Pareto IV Distribution}
First, we seek to describe the distribution of the number of hits to any of the $99$ websites, written $X_{\text{tot}}.$ A frequency table of $X_{\text{tot}}\mid X_{\text{tot}}\geq 1$ is displayed in the top of Figure \ref{fig:aggrHits} on a log-scale and reveals a surprisingly regular decay. At the value zero, however, the probability mass function forms an irregular peak : $88.1\%$ of hits are $0$ while only $2.4\%$ and $1.3\%$ are $1$ and $2$ respectively. Assuming that $X_{\text{tot}}\mid X_{\text{tot}}\geq 1$ follows a Poisson or a negative binomial distribution delivers a poor fit. We thus turn our attention to a more flexible discrete family distribution called the discrete Pareto IV ($\text{D-PIV}$) and defined by its probability mass function $p(k)= F(k)- F(k-1)$ on $\mathbb N=\{1,2,\ldots\},$ where 
$$ F(x)=1-\left\{1+\xi {(x+\mu)^{\beta}-\mu^{\beta} \over \sigma}\right\}^{-1/\xi} 1_{\{0\leq x<e_1\}},$$
for $\xi\in\mathbb R,$ $\mu\geq 0,$ $\beta>0,$ $\sigma>0,$ and, if $\xi\geq 0,$ $e_1=\infty.$ When $\xi>0,$ $F$ is the cumulative distribution function (cdf) of a Pareto IV distribution (see e.g.\ \citealp{arnold2015pareto}). When $\xi<0,$ $F$ has a finite endpoint 
$$e_1=\left(\mu^\beta+{\sigma\over |\xi|}\right)^{1/\beta}-\mu,$$
and when $\xi=0,$ it is extended by continuity to $F(x)=1-\exp\left\{ (x+\mu)^{\beta}/\sigma-\mu^{\beta} /\sigma\right\}.$ In the case $\mu=0,$ $\beta=1,$ the D-PIV coincides with a discrete generalized Pareto distribution which has been used, for instance, by \cite{Prieto2014} to model road accidents.


We fit the D-PIV to $X_{\text{tot}}\mid X_{\text{tot}}\geq 1$ by maximizing numerically the log-likelihood using the function $\texttt{optim}$ of $\texttt{R}$ with starting parameters $(\xi,\sigma,\beta,\mu)=(0.1,1,0.1,1)$ \citep{R2015}. Even though there are more than $200\, 000$ observations, the maximization is fast because the likelihood only needs to be evaluated once at each observed integer value. The D-PIV fits the data well as shown in the bottom of Figure \ref{fig:aggrHits}. This is confirmed by a discrete Kolmogorov--Smirnov goodness-of-fit test in package \texttt{dgof} applied to the test data set giving a p-value of $0.14$ \citep{Arnold2011}. Maximum likelihood estimates and $95\%$ confidence intervals based on asymptotic normality are displayed in the table below.
\begin{center}
\small
\begin{tabular}{cccc}
   $\xi$   & $\sigma$&    $\beta$ &  $\mu$  \\
   \hline
 $-0.17_{ \pm 0.03}$ &  $0.20_{  \pm 0.06}$ & $0.07_{ \pm 0.02} $&$ 1.13_{ \pm 0.04}$ \\
\end{tabular}
\end{center}

Since $\xi<0,$ the fitted distribution has a finite endpoint $e_1,$ The later is roughly $1.7$ times the largest observation in the data set which is $52\, 419$ hits to Facebook.

For each website $i=1,\ldots,d,$ we now fit to $X_i\mid X_i\geq 1$ a D-PIV and two embedded models corresponding to the cases $\mu=0$ and $(\mu=0,$ $\beta=1)$ and select among them according to the Bayesian information criteria (BIC). The following table shows the percentage of discrete Kolmogorov--Smirnov tests whose p-values are smaller than $0.05$ and the average sample size.


\begin{center}
\begin{tabular}{l|ccc}
    & Data Set &Rejected &  Av.\ Sample Size  \\
\hline 
D-PIV   & training &  $5.0\%$ & $2090$ \\
D-PIV   &  test &  $6.1\%$ & $234$ \\
\end{tabular}
\end{center}

The rejection rate is close to $5\%,$ indicating a good fit which supports the adequacy of the D-PIV family distribution to model $X_i\mid X_i\geq 1.$ The saturated model and the embedded models in the case $\mu=0$ and $(\mu=0,$ $\beta=1)$ were selected $25,$ $59$ and $15$ times respectively.

\section{The Censored Student Copula Graphical Model}

We are now interested in modeling the dependence structure of $\g X=(X_1,\ldots,X_d).$ The problem cannot easily be reduced to lower dimension by assuming independence between blocks of marginals because the dependence between them is rather strong. To illustrate this, we define a binary vector $\g B$ by $B_i=1_{\{X_i\geq 1\}},$ where $1_A$ is the indicator function, and perform $\chi^2$ tests of independence with level $0.05$ for all pairs; the test rejects independence $98\%$ of the time. Graphical models provide a way to simplify a joint density by making assumptions of conditional independence between marginals as we will discuss later. We use \texttt{ci.test} of package \texttt{bnlearn} to test conditional independence $B_i\ci B_j\mid \g B_{-i,-j}$ \citep{Scutari2010}. Pairwise conditional independence is rejected only $33\%$ of the time, suggesting that there might be many conditional independence relations to exploit to reduce dimensionality.

Is there a simple multivariate distribution that can be chosen for $\g X$? There are extensions of the Pareto IV in the multivariate case but they typically have limited dependence structure or are intractable (\citealp{arnold2015pareto}). As explained by copula theory, the marginals of any multivariate continuous distribution can be transformed to obtain another distribution with marginals of arbitrary continuous distribution while preserving the dependence structure (see e.g.\ \citealp{Sklar1959} or \citealp{Nelsen2007}). In the discrete case, copulas suffer from limitations such as non-identifiability \citep{Genest2007}.

An additional difficulty for modeling $\g X$, as we have seen, is that its distribution is fundamentally different on zero than on positive integers, requiring a discrete multivariate distribution that accounts for this mixture. The multivariate zero-inflated Poisson distribution is such an instance \citep{Li1999,Liu2015}. 

The multivariate Gaussian and Student distributions are some of the few known multivariate distributions with explicit multivariate marginal and conditional densities. Their dependence structure is well understood and has some flexibility without being overdetermined as it only involves pairwise interactions. We now explain how they can be used for modeling $\g X.$ Let $\g t\in \mathbb R^d$ be a vector of thresholds and suppose that $\g Z$ follows a centered Student distribution with degree of freedom $\nu\in (2,\infty]$ and covariance matrix $\frac{\nu}{\nu-2}\Sigma$ such that $\Sigma$ has diagonal $\g 1\in\mathbb R^d.$ By convention, $\g Z\sim \mathcal N(0,\Sigma)$ in the case $\nu=\infty.$ We obtain a vector $\g X$ with values in $(\{0\}\times \mathbb N_+)^d$ satisfying $X_i\mid X_i\geq 1\sim F_{\text{D-PIV}}$ for all $i$ by applying to $\g Z$ the following procedure: censor $Z_i$ when it falls below $t_i,$ set censored values to $0,$ transform non-censored $Z_i$ appropriately and round them. More precisely, define $\g X$ by 
\begin{align*}
X_i & =  \begin{cases}
\lfloor Y_i\rfloor + 1 & \text{if } Z_i\geq t_i,\\
0 & \text{else,}
\end{cases}
\qquad Y_i := F_{\text{PIV}_i}^{-1}\{ \Phi_{Z\mid Z\geq t_i}(Z_i\mid Z_i\geq t_i)\}, 
\end{align*}
for all $i=1,\ldots,d,$ where $\Phi_{Z\mid Z\geq t_i}$ is the cdf of a truncated Student and $F_{\text{PIV}_i}$ is such that $F_{\text{PIV}_i}(k)-F_{\text{PIV}_i}(k-1)$ is the probability mass function of the D-PIV fitted to $X_i\mid X_i\geq 1.$ Here, $\lfloor x\rfloor$ denotes the largest integer smaller or equal to $x.$

 This formulation provides a simple yet non trivial probability distribution with values in $\mathbb N^d$ accounting for the irregularity of the distribution between null and positive values. It has been frequently used to model multivariate rainfall data where zeros occur when no rain is measured, see e.g.\ \cite{bell1987space} or \cite{allcroft2003latent}. It also appears in multivariate extreme value analysis where zeros correspond to non-extremal events, such as a modeling of extremal oceanographic data in \cite{bortot2000multivariate}.

 In theory, we could express the probability mass function of $\g X$ as a sum of terms involving the joint cdf of $\g Z,$ but the expression is intractable when $d$ is large. We are thus treating positive values of $\g X$ as continuous, making the working assumption that $X_i\approx  Y_i+\frac{1}{2}$ when $Z_i\geq t_i.$ A similar approximation for the multivariate normal copula in the case of discrete data is studied in \cite{nikoloulopoulos2016efficient}. This is a reasonable assumption here because the range of positive values is relatively large. Another benefit is that it yields a tractable expression for conditional distributions: for any disjoint sets $A,B,C,D\subseteq \{1,\ldots,d\},$ $\g k\in \mathbb N^d_+,$
\begin{align}\label{eq:conditional}
\pr\left\{\g X_{AB}=(\g k_A,\g 0)\mid \g X_{CD}=(\g k_C, \g 0)\right\} & =
{ \Phi\left(\g t_{BD} \mid \g Z_{AC}=\g z_{AC}\right) \over \Phi \left( \g t_B \mid \g Z_{AC}=\g z_{AC}\right) }\, \phi(\g z_A\mid \g z_C),
\end{align}
where $z_i =  \Phi_{Z_i\mid Z_i\geq t_i}^{-1} \{ F_{\text{PIV}}(\g k_i-\frac{1}{2})  \}.$ The quantities $\Phi(\cdot\mid \g z_F)$ and $\phi(\cdot\mid \g z_F)$ denotes respectively the joint cdf and density function of $\g Z_E\mid \g Z_F=\g z_F$ for some disjoint sets $F,E\subset \{1,\ldots,d\}$ and coincides with the joint cdf $\Phi(\cdot)$ and density $\phi(\cdot)$ of a non-centered Student distribution on $\mathbb R^{|E|}$ \citep{ding2016conditional}. 
The joint cdf $\Phi(\cdot)$ can be evaluated using the package \texttt{mvtnorm} \citep{genz2008mvtnorm}. Notice that one can sample from $\g Z_E\mid \g Z_F$ using (\ref{eq:conditional}), for instance, by applying inverse transform sampling recursively.


We refer to the model above as the censored Student copula model and we now present a possible way of estimating it. Recall that the website visits $\g x=\{\g x^{(k)}\}_{k=1}^n$ are assumed to be sampled from $\g X.$ The parameters $t_i$ can directly be estimated as $\hat t_i=\Phi^{-1}(n_i/n)$ where $n_i$ is the number of observations such that $x^{(k)}_i=0.$ It remains to estimate the matrix $\Sigma$ to fully determine the distribution. Its maximum likelihood estimate in the Gaussian case would be
$$\hat S= {1\over n}\sum_{k=1}^n \g z^{(k)}  {\g z^{(k)}}^T,$$
 if a sample $\g z$ from $\g Z$ was observed. However, only a censored transformation of $\g z$ is observed and different methods are possible in this case (see e.g.\ \citealp{lee2012algorithms} or \citealp{Schemper2013}). We found satisfactory performance and efficiency of estimating $\rho=\Sigma_{ij}$ by maximizing the pairwise likelihood $\ell_{ij}(\rho)= \sum_{k=1}^n \log \pr(X_i=x^{(k)}_i,X_j=x^{(k)}_j)$ for all pairs $(i,j)$ separately, where
 \begin{equation*}
\begin{aligned}
\pr(X_i=0,X_j=0) &\; =\; \pr(Z_i\leq t_i,Z_j\leq t_j),\\
\pr(X_i=x_i,X_j=x_j) & \;=\; c_i(x_i) c_j(x_j) \,   \pr(Z_i= z_i,Z_j= z_j), \\
\pr(X_i=x_i,X_j=0)& \;=\; c_i(x_i)  \,  \pr(Z_i = z_i,Z_j\leq t_j), \\
z_i  \;=\; M_i^{-1}\{F_{X_i\mid X_i\geq 1}(x_i)\}, \quad  M_{i}(z)&\;=\;\pr(Z_i\leq z\mid Z_i\geq t_i), \quad c_i(x_i)   \;=\; { f_{X_i\mid X_i\geq 1}(x_i) \over M_i'(z_i)},
\end{aligned}
\end{equation*}
 where we abuse notation and write $\pr(Z_i= z):={\partial \over \partial z} \pr(Z_i\leq z).$ All together, this gives a consistent estimator $\hat \Sigma^C$ of $\Sigma$ defined by $\hat \Sigma^C_{ij}=\argmax_{\rho\in [-1,1]} \ell_{ij}(\rho).$

We compute $\Sigma^C$ in the case $\nu=\infty$ for $n=19\,436$ users and display the largest positive and negative correlations $\Sigma^C_{ij}$ in the table below with a $95\%$ confidence interval based on asymptotic normality of the maximum likelihood estimator.

\begin{center}
\small
\begin{tabular}{llll}
 i & j & $\rho_{ij}$ \\ 
 \hline
TurboTax & Intuit & $0.91_{ \pm 0.01}$ \\ 
 YouTube & YouTube-NoCookie& $0.86_{ \pm 0.00}$ \\ 
 YouTube & Vevo & $0.78_{ \pm 0.01}$ \\ 
 Expedia & TripAdvisor & $0.75_{ \pm 0.02}$ \\ 
 YouTube-NoCookie & Vevo & $0.74_{ \pm 0.01}$ \\ 
 Lowe's & The Home Depot & $0.71_{ \pm 0.02}$ \\ 
 eBay & PayPal & $0.68_{ \pm 0.02}$ \\ 
 Facebook & Zynga & $0.68_{ \pm 0.02}$ \\ 
 Blogger & WordPress & $0.67_{ \pm 0.02}$ \\ 
 U.S. Internal Revenue Service & TurboTax & $0.65_{ \pm 0.03}$ \\ 
\hline 
 American Express & Vevo & $-0.11_{ \pm 0.06}$ \\ 
 Comcast Digital Entertainment & Road Runner & $-0.10_{ \pm 0.05}$ \\ 
 Vevo & Discover & $-0.08_{ \pm 0.06}$\\
 \end{tabular}
\end{center}
$ $

The marked dependence between the websites above is sensible: Turbo Tax and Intuit are tax softwares, YouTube and Vevo video websites, Expedia and TripAdvisor travel websites, Lowe's and The Home Depot home improvement stores, eBay is an e-commerce company who owned PayPal, an online payment operator. Facebook, a social networking service, had a partnership with Zynga, an online video game provider; Blogger and WordPress are blog publishing services.

 Although there are many other positively correlated pairs of websites, only $3$ pairs exhibit a significant negative correlation according to our model. Visitors of Vevo are less interested in American Express and Discover, two financial service companies, as are visitors of Road Runner, a running shoe store, in Comcast, a mass media company.

Two characteristics are available on users: their age, which takes tabulated values and is censored above $65$, and their gender, which is binary. These covariates could be included in the analysis by regressing the parameters of the model with respect to them. We prefer here to treat them on the same terms as the other variables in a fully multivariate analysis. Let $Z_a$ and $Z_b$ be Student distributed. For age, we assume that $X_{a}= 1_{\{Z_{a}\geq t_{a}\}}$ for some $t_a\in\mathbb R.$ For gender, we suppose that $X_{b}$ is censored from above when $Z_{b}\geq t_{b}$ for some $t_b$ and $X_{b}= F^{-1}_{b}\{\Phi_{Z_{b}\mid Z_{b}\leq t_{b}}(Z_{b} \mid Z_{b}\leq t_{b})\},$ where $F^{-1}_{b}$ is the empirical quantile function of $\g x_b.$ Note that since $\g x_b$ is discrete but treated as continuous, it is a better approximation to transform it to uniform using $\{F_b(x)+F_b(x-1)\}/2$ instead of its empirical cdf $F_b.$ Following the procedure explained previously, we obtain an estimate for the threshold vector $\g t \in \mathbb R^{d+2}$ and the matrix $\Sigma\in\mathbb R^{d+2}\times \mathbb R^{d+2}.$ The underlying random vector is now $\g Z=(Z_1,\ldots,Z_{99},Z_a,Z_b)$ which is Student distributed with degree of freedom $\nu>2,$ mean $\g 0$ and covariance matrix $\frac{\nu}{\nu-2}\Sigma$.

The next table shows with which websites age and gender share the strongest correlation in the case $\nu=\infty$, with the convention $X_a=1$ for female and $X_a=0$ for male. 

\begin{center}
\small
\begin{tabular}{llll}
 i & j & $\Sigma^C_{ij}$ \\
 \hline 
  age & Legacy & $0.34_{ \pm 0.03}$ \\ 
 age & American Express & $0.29_{ \pm 0.04}$ \\ 
 age & WhitePages  & $0.28_{ \pm 0.03}$ \\ 
 age & Shopzilla & $0.27_{ \pm 0.03}$ \\ 
 age & SuperPages & $0.27_{ \pm 0.03}$ \\ 
\hline 
 age & Vevo & $-0.30_{ \pm 0.02}$ \\ 
 age & YouTube-NoCookie & $-0.23_{ \pm 0.02}$ \\ 
 age & Disney DOL & $-0.22_{ \pm 0.03}$ \\ 
 age & YouTube & $-0.21_{ \pm 0.02}$ \\ 
 age & Tumblr & $-0.20_{ \pm 0.03}$ \\ 
\end{tabular} 
\quad \quad
\begin{tabular}{llll}
 i & j & $\Sigma^C_{ij}$ \\
 \hline
 gender & Pinterest & $0.31_{ \pm 0.04}$ \\ 
 gender & JCPenney & $0.30_{ \pm 0.04}$ \\ 
 gender & Macy's & $0.29_{ \pm 0.04}$ \\ 
 gender & Allrecipes & $0.28_{ \pm 0.03}$ \\ 
 gender & Everyday Health & $0.27_{ \pm 0.03}$ \\ 
\hline 
 gender & ESPN  & $-0.25_{ \pm 0.03}$ \\ 
 gender & Turner-SI & $-0.22_{ \pm 0.03}$ \\ 
 gender & Big Lead Sports & $-0.21_{ \pm 0.04}$ \\ 
 gender & NFL & $-0.21_{ \pm 0.03}$ \\ 
 gender & CNET & $-0.17_{ \pm 0.04}$ \\ 
\end{tabular}
\end{center}
$ $

The most visited websites by older persons were Legacy, an online memorial provider, American Express, a financial service company, White Pages, a contact information provider, Shopzilla, a shopping website and Experian, a global information services group. Younger persons mostly hit to Vevo, YouTube, Disney, a mass media and entertainment company, and Tumblr, a social networking website.

Women were more frequently on Pinterest, a photo sharing website, JCPenney and Macy's, two department store chains, Allrecipes, a social networking service focused on food, and Everyday Health, a company producing content on wellness. Men, on the other hand, preferred sport related networks and CNET, a technology news website.

From a model selection point of view, it is natural to try to represent the matrix $\Sigma$ with fewer parameters. As we have seen, the marginals are strongly dependent so $\Sigma$ is unlikely to be sparse. Following the ideas of graphical models, one can try to exploit sparsity of the inverse matrix $\Sigma^{-1}$ which translates, in the Gaussian case, into conditional independence relations between some of the marginals: $\Sigma^{-1}_{ij}=0$ if and only if $Z_i\ci Z_j\mid \g Z_{\{1,\ldots,d\}\setminus\{i,j\}}$ (see e.g.\ \citealp{Lauritzen}). The constraint is a little different for the Student distribution: $\Sigma^{-1}_{ij}=0$ implies that $Z_i$ and $Z_j$ are conditionally uncorrelated given the rest of the vector \citep{finegold2011robust}. Besides reducing the number of parameters, these assumptions allow us to visualize the dependence structure as a graph $\mathcal G=(V,E)$ defined as follows: its set of nodes $V$ is $\{1,\ldots, d\},$ i.e., each node corresponds to a marginal of $\g Z,$ and its set of edges $E$ satisfies
\begin{align}\label{eq:defGM}
(i,j)\notin E  \;\implies \; \Sigma_{ij}^{-1}=0.
\end{align}

An efficient procedure to estimate a sparse matrix $\Sigma^{-1}$ given a sample $\g z$ i.i.d.\ drawn from $\mathcal N(0,\Sigma)$ is proposed in \cite{Friedman2008}. It solves the convex optimization problem called Gaussian graphical lasso which consists in maximizing the log-likelihood of a multivariate Gaussian with an additional penalization to enforce sparsity. More precisely, 
\begin{align}\label{eq:glasso} \hat \Theta =\underset{\Theta \succeq 0}{\argmax} \quad   \log \det\Theta-\operatorname{tr}(\hat S_n\Theta) -\lambda_n ||\Theta||_{1},\end{align}
where $\hat S_n=n^{-1} \g z^T \g z$ is the empirical covariance matrix, $\lambda_n>0$ is a regularization parameter, ``tr'' denotes the trace 
of a matrix, ``$\succeq 0$'' means non-negative definite and $||\Theta||_1:=\sum_{ij} |\Theta_{ij}|.$ After slightly reformulating (\ref{eq:glasso}), \cite{ravikumar2011high} show that for any consistent estimator $\hat S_n$ of $\Sigma$ (such as $\hat \Sigma^C$) and under an additional assumption (called incoherence or irrepresentability condition), the solution $\hat \Theta$ is a consistent estimator of $\Sigma^{-1}$ and it correctly detects null entries as $n\rightarrow \infty$ for fixed $d.$ When $\nu<\infty,$ solving (\ref{eq:glasso}) still makes sense as $\hat \Theta$ is the closest matrix to $\Sigma^{-1}$ in terms of a Bregman divergence. Note that if one is interested in estimating the graph $\mathcal G$ only, \cite{meinshausen2006high} present a procedure to estimate it consistently in the Gaussian case. There is a rich literature on Gaussian graphical lasso (see e.g.\ \citealp{banerjee2008model,yuan2010high,cai2016estimating}), copula Gaussian graphical models \citep{dobra2011copula,xue2012regularized,liu2012high}, Student graphical lasso \citep{finegold2011robust} and Gaussian lasso in the censored case \citep{johnson2009lasso}, including several variants such as the adaptive lasso which penalizes coefficients differently \citep{zou2006adaptive} and a decomposition of $\Sigma^{-1}$ into the sum of a sparse and a low-rank matrix \citep{chandrasekaran2010latent}.

Coming back to our data analysis, we compute $\hat S_n=\hat \Sigma^C$ for various degrees of freedom $\nu$ and solve (\ref{eq:glasso}) for several regularization parameters $\lambda$ using the package $\texttt{huge}$ \citep{zhao2012huge} to obtain matrices $\hat \Sigma$ whose inverse are sparse. We then choose $\nu$ and $\lambda$ by minimizing two scores: the average negative log-likelihood 
\begin{align}\label{eq:nllCensGM}
\ell(\hat \Sigma) =-\frac{1}{n} \sum_{k=1}^n \left\{ \log \Phi\left(\g t_{A_k^c}\mid  \g z_{A_k}^{(k)}\right) + \log \phi\left(\g z_{A_k}^{(k)}\right)+ \sum_{i\in A_k} \log c_i\left(x_i^{(k)}\right) \right\},
\end{align}
where $A_k=\{i:x^{(k)}_i\neq 0\},$ and the sum of average pairwise negative log-likelihood
$$\ell_{\text{pairwise}}(\hat \Sigma) =  \sum_{i<j} \ell_{ij}(\hat \Sigma_{ij}).$$
 We compute $\ell$ and $\ell_{\text{pairwise}}$ from the test data set using respectively $400$ and $19\, 436$ observations and display them in the table below. We also report the percentage of null entries in the upper triangular part of $\hat \Sigma^{-1}$ referred to as sparsity.

\begin{center}
\small
\begin{tabular}{ll|lllll}
& &  $\hat \Sigma^C$ &  $\hat \Sigma_{\lambda= 0.0075}$ &$\hat \Sigma_{\lambda= 0.02}$ & $\hat \Sigma_{\lambda= 0.1}$  \\
\hline
$\nu=\infty$& sparsity & $0\%$ &  $25\%$ & $50\%$ & $73\%$  \\
& $\ell_{\text{pairwise}}$ &   3079.33  &   \textbf{3079.31}  & 3079.36&    3081.78\\    
& $\ell$  & 23.14 & \textbf{23.07} & 23.09&  23.43 \\
\hline
$\nu=60$& sparsity & $0\%$ &     $24\%$     & $50\%$     & $75\%$  \\
&$\ell_{\text{pairwise}}$ & 3079.79 & 3079.78  & 3079.83& 3082.15 \\
& $\ell$ &23.20& 23.12 & 23.11 &23.43  \\
 \hline
 $\nu=30$&sparsity & $0\%$ &  & & & \\
&$\ell_{\text{pairwise}}$ &3080.19  & & & \\
& $\ell$ &23.24  & & & \\
\end{tabular}
\end{center}
$ $

\begin{figure}
\begin{center}
    \begin{subfigure}[b]{1.08\textwidth}
        \includegraphics[width=\textwidth, trim= 2cm 0.5cm 0cm 1cm]{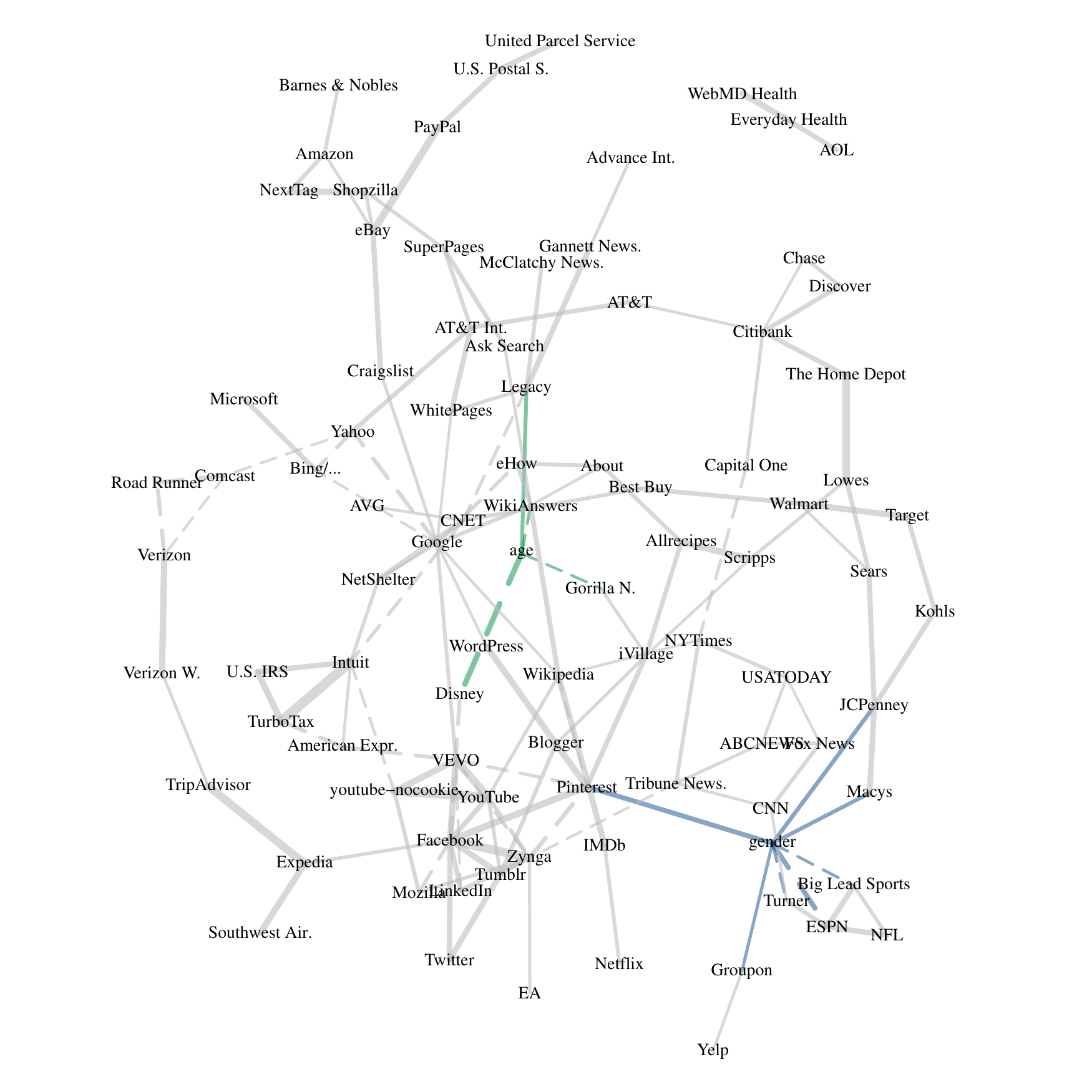}
    \end{subfigure}
        \caption[Graph illustrating the dependence between website visits]{Graph illustrating the dependence structure of $\g X=(X_1,\ldots,X_{99},X_a,X_b),$ the random vector of hits to the $99$ most visited websites; $X_a$ is the age and $X_b$ the gender of a user. We fitted a censored Gaussian copula graphical model to $\g X$ with sparse inverse covariance matrix $\Sigma^{-1}.$ The width of an edge $(i,j)$ in the graph is proportional to the interaction coefficient $\Sigma^{-1}_{ij}$ on a log-scale and shows how predictive $X_i$ is for $X_j$. We displayed only edges having the $5\%$ largest positive or negative coefficients and non-isolated nodes\footnotemark. Dashed edges correspond to positive coefficients, i.e., negative partial correlations. }
    \label{fig:graph}
    \end{center}
\end{figure}

Both scores agree on selecting the censored Gaussian copula graphical model with $\lambda=0.0075,$ corresponding to a sparsity of $25\%.$ Figure \ref{fig:graph} shows the corresponding graph, plotting only edges $(i,j)$ such that $\hat \Sigma^{-1}_{ij}$ are among the $5\%$ largest non-null entries in absolute values. An edge between $i$ and $j$ should be interpreted as $X_i$ carrying relevant information to predict $X_j.$ We identify several meaningful clusters such as financial companies (Chase, Discover, Citibank, Capital One), news (Fox News, CNN, ABCNEWS, USATODAY, NYTimes, Tribune Newspaper) and sports (Turner-SI, NFL Internet Network, Big Lead Sports, ESPN). The graph also illustrates which websites a user tends to visit depending on its age and gender and we recognize some of the relations discussed earlier.

\footnotetext{The following websites do not have any neighbor in the graph in Figure \ref{fig:graph}: Apple, Glam Media, Weather Channel Network, Bank of America, Adobe, Wells Fargo, NBC Universal Sites, Napster, ShopAtHome, Hewlett Packard, Experian.}

\begin{figure} 
    \centering
    \begin{subfigure}{\textwidth}
        \includegraphics[width=\textwidth, trim=0cm -1cm 0cm 0cm]{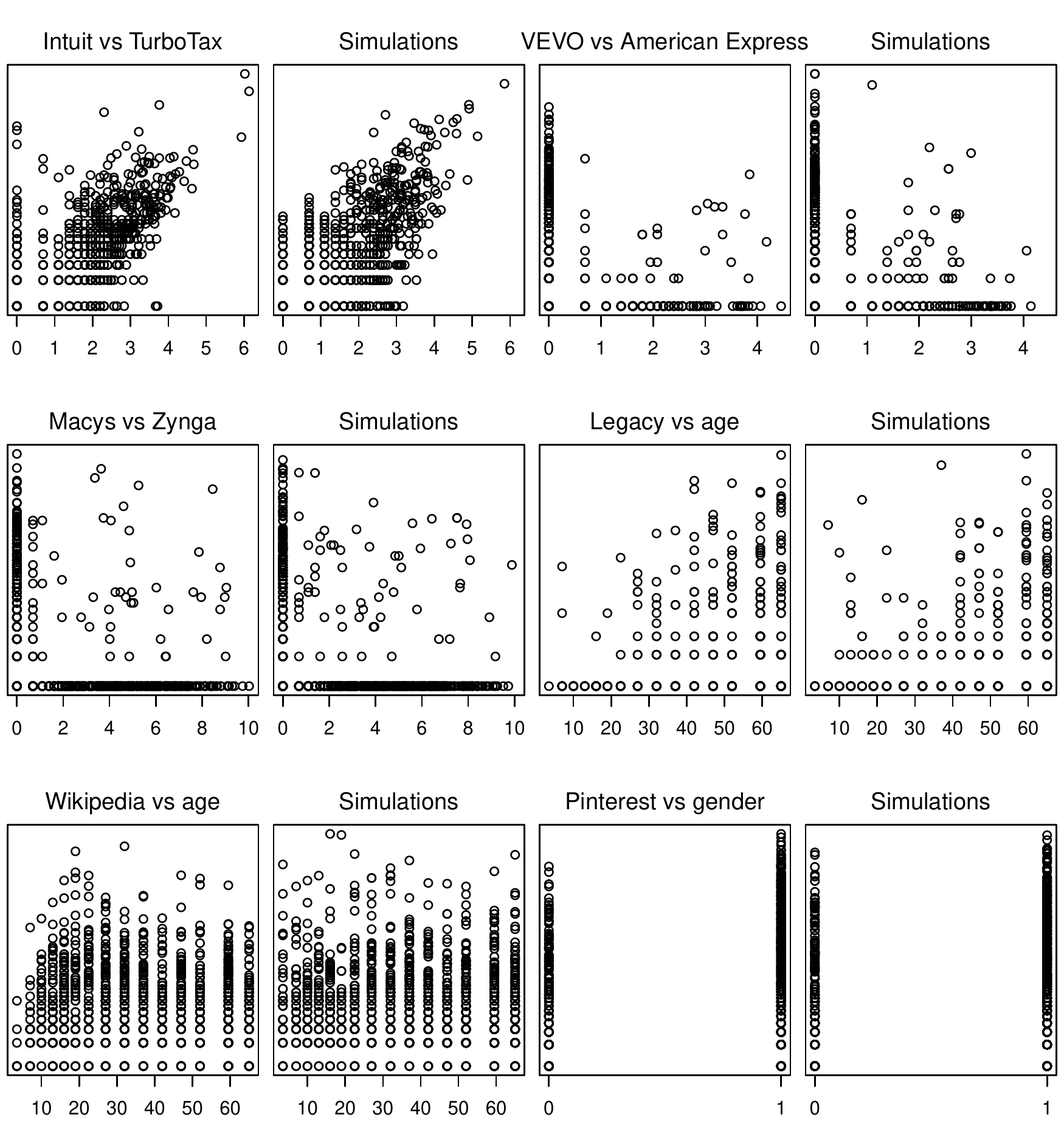}
    \end{subfigure}
        \caption[Scatterplots of website visits and simulations]{Several scatterplots of the data (first and third columns) and simulations from a censored Gaussian copula graphical model (second and fourth columns). The quantities plotted are the age $X_a,$ the gender $X_b$ and a transformation $\log(X_i+1)$ of the number of hits $X_i$ to website $i.$}
    \label{fig:scatterplots}
\end{figure}

\begin{figure} 
\begin{center}
    \begin{subfigure}[b]{0.45\textwidth} 
        \includegraphics[width=\textwidth, trim=1cm 1cm 1cm 1cm]{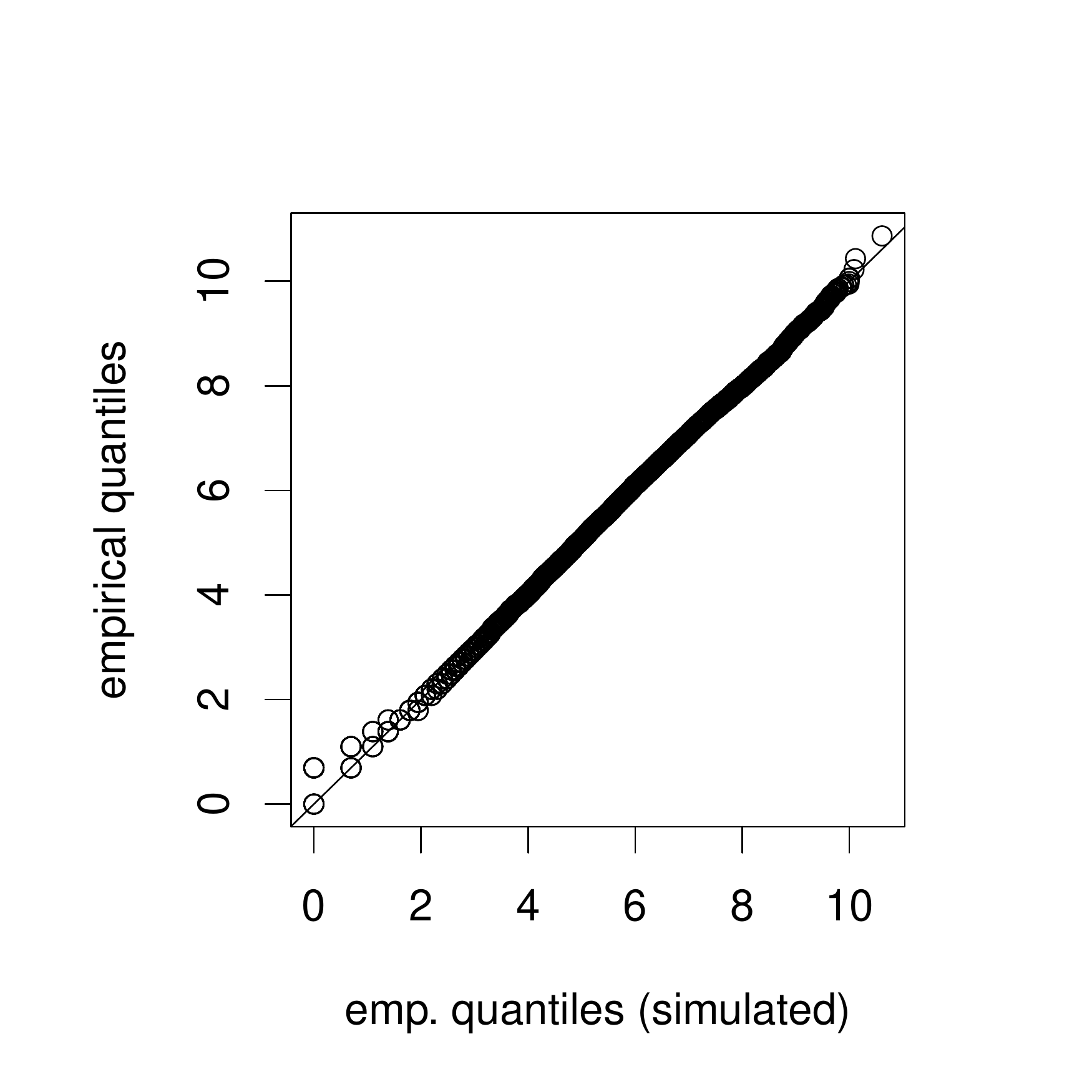}
    \end{subfigure}
    \begin{subfigure}[b]{0.45\textwidth} 
        \includegraphics[width=\textwidth, trim=1cm 1cm 1cm 1cm]{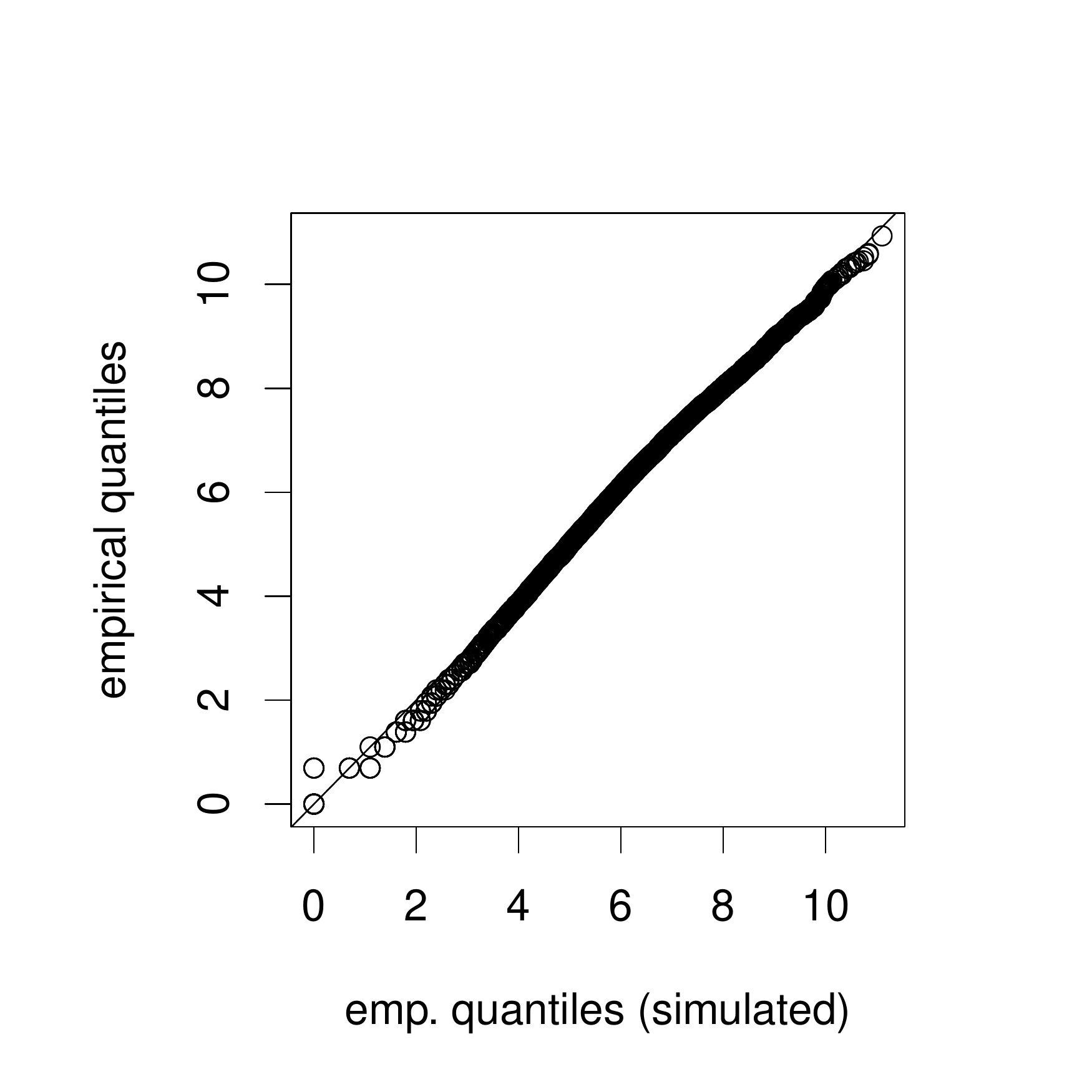} 
    \end{subfigure}
        \caption[Diagnostic QQ-plots]{Diagnostic QQ-plots for the sum of hits to the $5$ (left) and $99$ (right) most visited websites. Empirical quantiles of the data are plotted against simulated data from the censored Gaussian copula graphical model on a log scale.}
    \label{fig:qqplotsSum}
    \end{center}
\end{figure}

We now present several diagnostics of the selected model.  We start by testing if pairs from the real data set have the probability expected from the model of occurring in certain regions. Binomial tests with level $0.05$ are performed for every pairs $(i,j)$ and the result is reported in the next table.

\begin{center}
\small
\begin{tabular}{c|c}
Region &  Rejected  \\
\hline
$\pr(X_i>0,X_j>0)$ & $10.4\%$\\
$\pr(X_i=0,X_j>0)$& $7.7\%$ \\
\end{tabular}
\end{center}
$ $

The percentage of rejection is above $5\%,$ the rate of a correct model.

As a further diagnostic, we draw $19\, 436$ realizations from the censored Gaussian copula graphical model and compare them to the data. Figure \ref{fig:scatterplots} displays a few scatterplots chosen as follows: most positively, most negatively and least correlated pairs of website hits; most positively and less correlated pairs involving age; most correlated pair involving gender. Replicates from the model appear to be relatively similar to the data. As another comparison, Figure \ref{fig:qqplotsSum} shows QQ-plots for the sum of hits to the $5$ and $99$ most visited websites. The quantiles form a slightly concave line instead of a straight line, revealing some inaccuracies.

We end by assessing the performance of the model throughout the following experiment. We try to predict if $X_i>0$ given $\g X_{-i}.$ For $i=1,\ldots,99,$ this corresponds to predicting if a user visits website $i$ given its age, gender and visits on the other websites during the month. For gender, it corresponds to predicting if the user is a female. For age, we try to predict if $X_a>35,$ i.e., if the user is older than $35.$ To gain efficiency, we make the working assumption that $X_i\ci X_j\mid \g X_{\{1,\ldots,d\}\setminus\{i,j\}}$ if $\Sigma_{ij}^{-1}=0,$ a relation which holds for $\g Z$ but only approximately for $\g X.$ We choose the score function 
$$1_{\{X_i>0\}} \log \hat p+(1-1_{\{X_i>0\}})\log(1-\hat p),$$ where $\hat p$ is the predicted probability that $X_i>0,$ and compare the predictions of the model to various methods. First, we naively approximate $\pr(X_i>0\mid \g X_{-i})$ by $\pr(X_i>0)$. Second, we fit an Ising model --- a probabilistic graphical model for binary data --- to $\g B=(1_{X_1>0},\ldots,1_{X_{d+2}>0})$ using the \texttt{R} package \texttt{IsingFit} and compute $\pr(X_i>0 \mid \g B_{-i})$ \citep{van2014package}. Third, we train decision tree algorithms found in packages \texttt{rpart}, \texttt{tree}, \texttt{ctree} and \texttt{randomForest} (using $50$ trees for the latter) to predict $X_i>0 \mid \g X_{-i},$  \citep{van2014package,therneau2015package,hothorn2006unbiased,liaw2002classification}. The next table shows the average score and the percentage of correct guessing computed for about $400$ predictions from the test data set.

$ $

\begin{center}
\center
\small
\begin{tabular}{lllllllllll}
 Naive &  $\hat \Sigma^C$ &  $\hat \Sigma_{\lambda= 0.0075}$ &$\hat \Sigma_{\lambda= 0.02}$ & $\hat \Sigma_{\lambda= 0.1}$ & Ising & \texttt{rpart} & \texttt{tree} & \texttt{ctree} & RF   \\
 \hline
      $0.334$  &     $\textbf{0.254} $   &   $0.255 $  &   $ \textbf{0.254}   $  & $0.257 $ &   $  0.267 $ &     $0.324 $   &   $0.293$  &     $0.287$ &        $\infty$\footnotemark\\
      $86.8\%$  &    $89.6\%$   &    $89.6\%$     &  $89.3\%$    &   $89.9\%$   &    $89.3\%$       & $88.5\%$  &     $88.5\%$    &   $88.1\%$  &    $89.8\%$
      \end{tabular}
      \end{center}

$ $      

$ $

\footnotetext{\texttt{RandomForest} returns $7$ infinite values; its average score computed from finite values only is $0.221$.}

The censored Gaussian copula is a simple probabilistic model that provides an approximation of the full joint distribution of the data and is thus valuable to comprehend its dependence structure as a whole. The experiment above suggests that decision trees that were specifically trained for certain predictions do not outperform the model, at least when they are used without further tuning. The Ising model performs relatively well and seems a suitable alternative if only the fact that a website is visited matters for the analysis, but not the number of hits. On average, the censored Gaussian gives the best performance, showing its ability to capture relevant information in this data set.

\paragraph*{Acknowledgements} We would like to thank Nielsen Holdings for providing the data, and the first author is grateful to the Berrow Foundation for financial support.

\newpage

\bibliography{references}
\end{document}